\documentclass[aip,apl,reprint]{revtex4-1}

\usepackage[utf8]{inputenc}
\usepackage{graphicx}
\usepackage{dcolumn}
\usepackage{bm}
\usepackage[dvipsnames]{xcolor}
\usepackage{hyperref}
\usepackage{siunitx}
\usepackage{mathrsfs}
\usepackage{braket}
\usepackage{acronym}
\usepackage{mathtools}
\usepackage{amsmath}
\usepackage{placeins} 

\hypersetup{colorlinks=true, linkcolor=blue, citecolor=blue, urlcolor=blue}

\makeatother

\begin{document}

\newacro{qd}[QD]{quantum dot}
\newacro{hom}[HOM]{Hong-Ou-Mandel}
\newacro{dbr}[DBR]{distributed Bragg reflector}
\newacro{apd}[APD]{avalanche photodiode}
\newacro{pbs}[PBS]{polarizing beam splitter}
\newacro{x}[X]{exciton}
\newacro{eom}[EOM]{electro-optic modulator}
\newacro{qwp}[QWP]{quarter-wave plate}
\newacro{hwp}[HWP]{half-wave plate}
\newacro{pll}[PLL]{phase-locked loop}
\newacro{hbt}[HBT]{Hanbury Brown and Twiss}
\newacro{er}[ER]{extinction ratio}
\newacro{snspd}[SNSPD]{superconducting nanowire single-photon detector}
\newacro{tpi}[TPI]{two-photon interference}
\newacro{spdc}[SPDC]{spontaneous parametric down conversion}


\title{Fast and efficient demultiplexing of single photons from a quantum dot with resonantly enhanced electro-optic modulators} 

\author{Julian M\"unzberg}
\email{julian.muenzberg@uibk.ac.at}
\author{Franz Draxl}
\affiliation{Institut f\"ur Experimentalphysik, Universit\"at Innsbruck, Technikerstr.~25, 6020 Innsbruck, Austria}

\author{Saimon Filipe Covre da Silva}
\affiliation{Institute of Semiconductor and Solid State Physics, Johannes Kepler University Linz, 4040 Linz, Austria}

\author{Yusuf Karli}
\affiliation{Institut f\"ur Experimentalphysik, Universit\"at Innsbruck, Technikerstr.~25, 6020 Innsbruck, Austria}

\author{Santanu Manna}
\author{Armando Rastelli}
\affiliation{Institute of Semiconductor and Solid State Physics, Johannes Kepler University Linz, 4040 Linz, Austria}

\author{Gregor Weihs}
\author{Robert Keil}
\affiliation{Institut f\"ur Experimentalphysik, Universit\"at Innsbruck, Technikerstr.~25, 6020 Innsbruck, Austria}

\date{\today}

\begin{abstract}
We report on a multi-photon source based on active demultiplexing of single photons emitted from a resonantly excited GaAs quantum dot. Active temporal-to-spatial mode demultipexing is implemented via resonantly enhanced free-space electro-optic modulators, making it possible to route individual photons at high switching rates of \SI{38}{MHz}. We demonstrate routing into four spatial modes with a high end-to-end efficiency of $\approx\SI{79}{\%}$ and measure a four-photon coincidence rate of \SI{0.17}{Hz} mostly limited by the single-photon source brightness and not by the efficiency of the demultiplexer itself. We use the demultiplexer to characterize the pairwise indistinguishability of consecutively emitted photons from the quantum dot with variable delay time.
\end{abstract}


\maketitle 

\section{\label{sec:introduction}Introduction}

The interference of multiple non-interacting, indistinguishable particles is the driving mechanism behind the rich dynamics of multi-photon quantum optics\cite{Hong1987, Tichy2014, Menssen2017, Jones2020, Muenzberg2021} and sets the stage for advanced quantum information protocols, such as boson sampling,\cite{Aaronson2011,Wang2017,Loredo2017,Wang2019,Zhong2020} quantum simulation,\cite{Aspuru2012, Sparrow2018} and quantum networks.\cite{Pompili2021}
The required multi-photon states are typically generated by \ac{spdc} sources.\cite{Shih2003,Christ2013} Via advanced source engineering\cite{Mosley2008,Pickston2021} or tight spectral filtering,\cite{Scheidl2014} these sources produce photons with almost perfect indistinguishability and degree of entanglement.\cite{Proietti2021} Nevertheless, due to the probabilistic nature of the \ac{spdc} process there is an inherent and unavoidable trade-off between the source brightness and an undesired background from higher-order emissions. 

This severe limitation can be avoided by using true single-photon emitters, such as epitaxial semiconductor \acp{qd} which, by virtue of the deterministic nature of the emission process, offer an alternative route towards on-demand generation of multi-photon states with precisely defined photon number.\cite{Senellart2017}
By collecting the emission of $n$ remote \acp{qd} pumped by the same excitation laser, an $n$-photon source can, in principle, be constructed. However, it has proven difficult to generate indistinguishable photons from separate \acp{qd} due to variations in the \acp{qd}' structural properties and noise in their solid-state environment. In order to obtain a source of multiple indistinguishable photons from remote \acp{qd}, the emission wavelengths of the emitted photons, as well as the spectral and temporal overlap of the wavepackets need to be exactly matched. Moreover, the noise in the solid-state environment needs to be eliminated. Nevertheless, just recently, \ac{tpi} between photons from two remote \acp{qd} with a visibility of 93\% was demonstrated.\cite{Zhai2021} 
Despite this result, it remains difficult to scale this method to a larger number of photons due to the required experimental overhead. Upscaling demands either expensive equipment (a cryostat per \ac{qd} single-photon source) or sophisticated optics to couple the emission of multiple \acp{qd} from a single sample into separate optical fibers in addition to electric control of individual \acp{qd}.

Alternatively, it is also possible to implement a multi-photon source by temporal-to-spatial mode demultiplexing, where $m$ consecutively emitted photons from a single \ac{qd} are routed into $m$ spatial modes. 
Probabilistic demultiplexing with passive beamsplitters only constitutes a non-scalable approach as the detected $m$-photon rate scales with $(1/m)^m$.\cite{Loredo2017} On the other hand, active demultiplexing, our method of choice, is a scalable approach. Here, a photon in a particular temporal mode is actively and deterministically routed into a particular spatial mode with e.g.~\acp{eom} and polarizing beamsplitters. 
Experimentally, active temporal-to-spatial demultiplexing was implemented in an integrated lithium niobate waveguide with four photons (with a fast switching rate of \SI{40}{MHz} but low overall efficiency, due to waveguide-induced losses),\cite{Lenzini2017} in low-loss free-space setups via broadband \acp{eom} (slow switching at $\approx\SI{1}{MHz}$) and polarizing beamsplitters with four\cite{Hummel2019a} and 20 photons\cite{Wang2019} (of which 14 photons were detected), as well as with an acousto-optic modulator (slow switching at $\approx\SI{1.4}{MHz}$).\cite{Pont2022}

We use in this work resonantly enhanced \acp{eom} that require much lower half-wave voltages compared to their broadband counterpart and allow for much faster switching rates. Therefore, our implementation sets itself apart from the previous ones by demonstrating high switching rates (switching of individual photons at \SI{38}{MHz}) and high efficiency at the same time. A high switching rate allows us to route photons one-by-one instead of in bursts which is advantageous in terms of indistinguishability and detector dead time (see Sec.~\ref{sec:conclusion} for more details).

As a single-photon source, we use GaAs \acp{qd} grown by droplet etching epitaxy,\cite{CovredaSilva2021} which emit in the range of 780-\SI{805}{nm} wavelength. They have excellent emission properties, among them a high single-photon purity,\cite{Schweickert2018} near zero fine structure splitting,\cite{Huo2013} and fast radiative decay rates, which makes these \acp{qd} appealing single-photon and entangled photon pair sources.\cite{Huber2018, CovredaSilva2021} 

In this work, we implement active temporal-to-spatial mode demultiplexing of photons from a GaAs \ac{qd} single-photon source into four spatial modes. Pulsed resonant excitation of the \ac{qd} allows us to generate a train of single photons in well defined temporal modes (ideally a single photon per excitation laser pulse). 
Subsequently, we actively route the train of single photons into four spatial modes with resonantly enhanced \acp{eom} and \acp{pbs}. Compared to other excitation schemes, resonant excitation yields the highest photon indistinguishability between consecutively emitted photons,\cite{Reindl2019, Scholl2019} which is beneficial for multi-photon interference experiments.

This work is structured as follows. In Sec.~\ref{sec:methods}, we describe the experimental setup for single-photon creation and routing. We present the performance of our \ac{qd} single-photon source, as well as the demultiplexer in Sec.~\ref{sec:results}, and, as a proof of concept of our device, measure the indistinguishability of the \ac{qd} emission at various time delays in Sec.~\ref{sec:hom}. Finally, we conclude in Sec.~\ref{sec:conclusion}.

\section{\label{sec:methods}Demultiplexing single photons from a quantum dot}

\begin{figure*}[t]
\includegraphics[width=\linewidth]{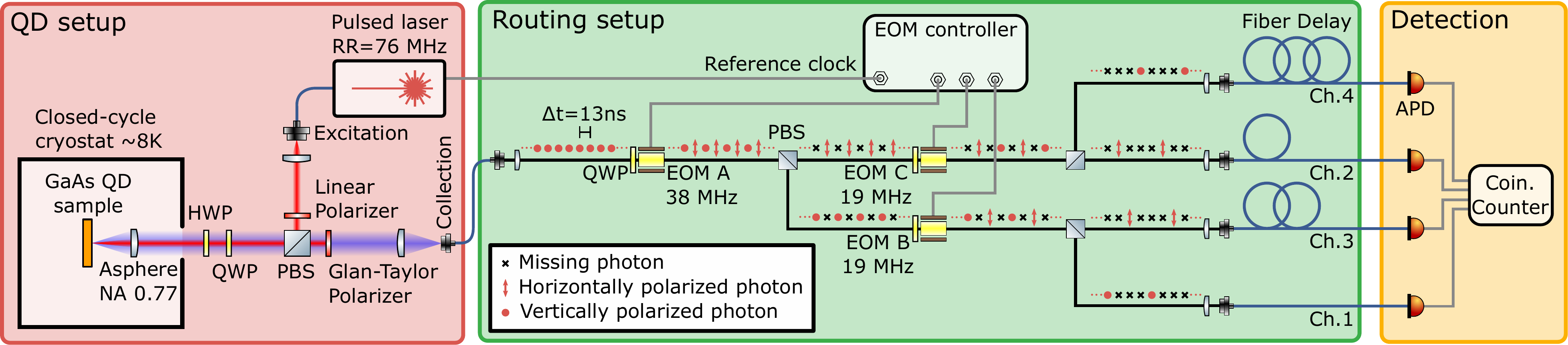}
\caption{\label{fig:setup}A schematic illustration of the experiment with the \ac{qd} setup (red box), the routing setup (green box), and the detection setup (yellow box). A detailed description of the setup is found in Section \ref{sec:methods}.}
\end{figure*}
Figure \ref{fig:setup} shows the experimental setup, which consists of three parts: 
Single photon generation in the \ac{qd} setup via resonance fluorescence (red box), active temporal-to-spatial mode demultiplexing in the routing setup (green box), and, lastly, single-photon detection and coincidence counting (yellow box).
Our GaAs \ac{qd} sample is grown with the local droplet etching technique by molecular beam epitaxy. The \ac{qd} layer is located in the center of a planar lambda cavity to enhance the extraction efficiency from the high refractive index material (for more details on the sample structure see Sec.~S1 in the \hyperlink{supp}{supplementary material}). From a finite difference time domain simulation of the structure, we retrieve an extraction efficiency of $\eta_\mathrm{extr}\approx 0.12$ (collected into an NA of $0.77$) and from a mode overlap calculation of the \ac{qd} emission with a single-mode fiber we obtain as fiber-coupling efficiency $\eta_\mathrm{fibercoup}\approx 0.60$.

The \ac{qd} sample is situated on a three-axis piezoelectric stage inside a closed-cycle cryostat with a base temperature of $\approx\SI{8}{K}$. We excite the \ac{qd} with a pulsed Ti:Sapphire laser (Coherent Mira 900) with a repetition rate $RR=\SI{76.2}{MHz}$ and a pulse duration of $\approx\SI{6}{ps}$ (after spectral filtering in a $4f$ pulse shaper). The excitation laser is focused onto the \ac{qd} with a cold aspheric lens ($\mathrm{NA}=0.77$) inside the cryostat so that
a stable coupling efficiency of the \ac{qd} emission to the single-mode fiber is maintained over days.

For laser rejection under resonant excitation, we rely on the so-called cross-polarized excitation/collection configuration.\cite{Kuhlmann2013} In our setup, the excitation laser passes a nanoparticle linear film polarizer (extinction ratio $>10^5:1$) in order to set the polarization to vertical before reflecting off a \ac{pbs} and propagating towards the \ac{qd} sample. Ideally, any back-reflected or back-scattered laser light is still vertically polarized and therefore blocked from transmitting into the collection path by the \ac{pbs} (extinction ratio $>10^3:1$) and an additional Glan-Taylor polarizer (extinction ratio $>10^5:1$). Thus, only the horizontally polarized emission from the \ac{qd} is transmitted and coupled into a polarization-maintaining single-mode fiber. We use a \ac{qwp} to compensate for small amounts of birefringence in the sample path which can cause a deviation from vertical polarization of any reflected laser light. A \ac{hwp} is used to set the excitation polarization of the \ac{qd}, which acts as a full-wave plate for the reflected laser, since it passes the waveplate twice. The two waveplates and the nanoparticle linear film polarizer are mounted on high-precision piezo-driven rotation stages with a minimal incremental motion of \SI{5}{\micro rad}, which is important to precisely optimize the laser rejection.

The output of the \ac{qd} setup is connected to the input of the routing setup. Here, the light is coupled out of the fiber, passes through a \ac{qwp} (see below for its role) and an \ac{eom} (QUBIG AM7R3-NIR-39). 
We apply a sinusoidal voltage to the \ac{eom} with a frequency equal to half the repetition rate of the pump laser $f_\mathrm{EOM\,A}=RR/2=\SI{38.1}{MHz}$ and an amplitude equal to half the half-wave voltage $V_\pi/2$ of the \ac{eom}. In resonantly enhanced \acp{eom}, the electro-optic crystal is embedded in a high-$Q$ resonant LC circuit which boosts the RF voltage across the crystal (the capacitor in the LC circuit) at the resonance frequency. Therefore, significantly less input voltage ($V_\pi\approx \SI{10}{V}$) is required compared to broadband \acp{eom} ($V_\pi> \SI{1}{kV}$). 
The applied sinusoidal voltage is phase locked to the pump laser via a reference clock input and a \ac{pll}. The \ac{qwp} oriented at \SI{45}{\degree} transforms the initially linear polarization to circular polarization, which is equivalent to biasing the \ac{eom} at $V_\pi/2$. Due to the biasing, a full modulation from vertical to horizontal polarization is achieved. If the phase of the applied sinusoidal voltage (with respect to the reference clock) is set such that a photon passes through the \ac{eom} at the maximum or minimum of the sinusoidal modulation, then every second photon will be switched from vertical to horizontal polarization. Subsequently, the photons are routed according to their polarization with a \ac{pbs}. Then, in the absence of losses, in transmission (reflection) every second time bin (duration $\Delta t=\SI{13.1}{ns}=RR^{-1}$) would be occupied by a horizontally (vertically) polarized photon, respectively, and all other time bins would be empty.
The non-linear transfer function (Malus' Law) of the \ac{eom}-\ac{pbs} system causes a flattening of the minima and maxima of the transmission (reflection) function in time, such that the adjustment of the relative phase to the pump laser and the half-wave voltage is uncritical (see Sec.~S2 in the \hyperlink{supp}{supplementary material} for more details).

In the second stage of the routing setup, we use two beam lines each with a \ac{qwp}, an \ac{eom} (QUBIG AM7R3-NIR-19) operated at a quarter of the pump laser repetition rate $f_\mathrm{EOM\,B}=f_\mathrm{EOM\,C}=RR/4=\SI{19.05}{MHz}$, and a \ac{pbs}
to route four photons in four consecutive time bins into four spatial modes labelled Ch.~1 to 4. Furthermore, we compensate the temporal delay between the photons by using different fiber lengths of \SI{2}{m}, \SI{4.7}{m}, \SI{7.4}{m}, and \SI{10.1}{m} for Ch.~1 to 4, respectively. In order to make the photons also indistinguishable in their polarization degree of freedom, the fiber couplers in Ch.~1 and 4 are rotated by $\SI{90}{\degree}$ with respect to the fiber couplers in Ch.~2 and 3, such that the light is always coupled into the fast axis of the polarization-maintaining single-mode fiber.

In order to measure the four-photon coincidence rate of our multi-photon source, we attach each output channel to an \ac{apd} and use a coincidence counter to measure the four-photon coincidence rate.

\section{\label{sec:results}Performance of the multi-photon source}

\begin{figure*}[t]
\includegraphics[scale=1]{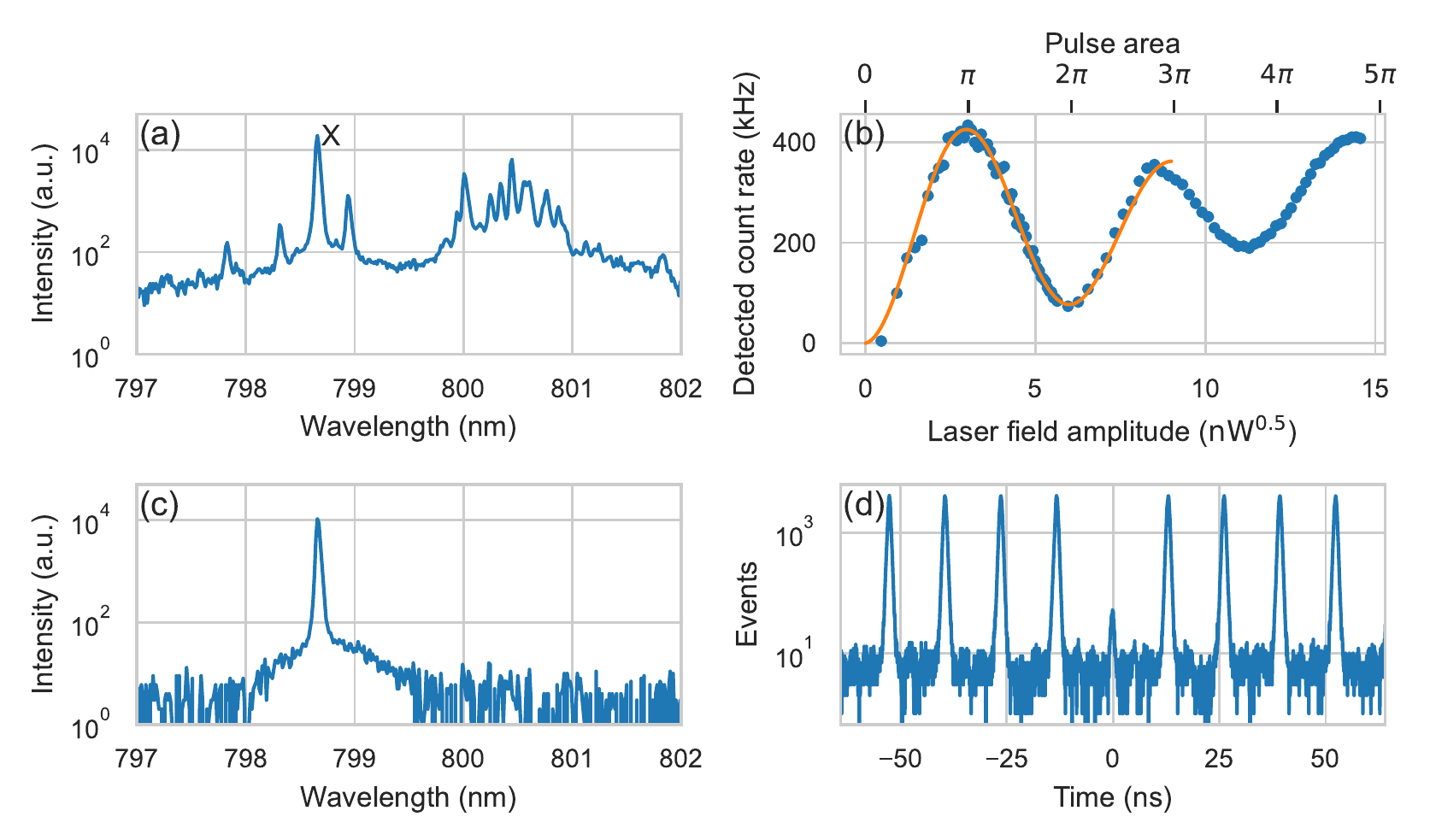}
\caption{\label{fig:qd_characterization}Characterization of the \ac{qd} emission. (a) Background-subtracted spectrum of the \ac{qd} under above bandgap excitation with laser light at \SI{532}{nm}. The neutral exciton (X) emission wavelength is \SI{798.66}{nm}. (b) Power-dependent Rabi oscillations of the neutral exciton up to a pulse area of approx.~$5\pi$. The detected single-photon count rate reaches a maximum of $\approx\SI{425(4)}{kHz}$ at a $\pi$-pulse power of \SI{9}{nW} (measured in front of the cryostat window, focused laser beam diameter $\approx\SI{1.5}{\micro m}$). For $\pi$-pulse excitation, we extract an excited state population probability of $90.9(5)\%$ from the fit (orange line). (c) Background-subtracted spectrum of the \ac{qd} under pulsed resonant s-shell excitation with a $\pi$-pulse. (d) Second-order intensity correlation histogram under pulsed resonant excitation with $g^{(2)}(0)=0.016(1)$.}
\end{figure*}
First, we characterize the performance of our \ac{qd} single-photon source in terms of brightness and single-photon purity. By turning off all \acp{eom} and removing the \acp{qwp} from the routing setup, ideally all light is routed towards Ch.~1 (compare Fig.~\ref{fig:setup}), which can be connected to a spectrometer or an \ac{apd} depending on the type of measurement. In this configuration of the routing setup, the routing efficiency into Ch.~1 is $\eta_\mathrm{Ch1}=0.91(4)$ (obtained from a separate measurement with laser light).

Figure \ref{fig:qd_characterization} shows the recorded data for the one \ac{qd} which is used as the single-photon source throughout this work. In Fig.~\ref{fig:qd_characterization}(a), the spectrum under above bandgap excitation with \SI{532}{nm} laser light is shown on a logarithmic scale. The neutral exciton transition corresponds to a wavelength of \SI{798.66}{nm}.
Figure \ref{fig:qd_characterization}(c) shows the spectrum under pulsed resonant s-shell excitation of the exciton. In this case, we tune the center wavelength of the pump laser exactly in resonance with the exciton transition, such that we only observe fluorescence from the pronounced zero-phonon line of the exciton, as well as a very weak phonon sideband (visible as pedestal of the main peak).
Next, we measure the \ac{qd} fluorescence signal as a function of the excitation laser power. The signal is monitored with an \ac{apd} (detection efficiency $\eta_\mathrm{det}\approx 68(5)\%$) without any spectral filtering of the signal. The result is plotted in Fig.~\ref{fig:qd_characterization}(b), which shows a clear modulation of the signal indicative of Rabi oscillations up to a pulse area of $\approx 5\pi$ with a maximum detected count rate of $\SI{425(4)}{kHz}$ at a pulse area of $\pi$. By fitting the data 
(see Sec.~S3 in the \hyperlink{supp}{supplementary material}) we retrieve a population probability $\eta_\mathrm{pop}$ of the neutral exciton of $90.9(5)\%$. Above a pulse area of $3\pi$, the signal deviates from the damped Rabi-oscillation-like behavior most probably due to laser leakage and off-resonant excitation of additional emission lines. The latter is confirmed by recording the spectrum for excitation powers $>3\pi$. The spectrum shown in Fig.~\ref{fig:qd_characterization}(c) is recorded at $\pi$-pulse excitation. 

Also at $\pi$-pulse excitation, we measure the single-photon purity in a \ac{hbt} measurement. We perform the measurement by again inserting the \ac{qwp} before \ac{eom} A such that half the light is probabilistically guided towards output Ch.~2 and half the light towards output Ch.~1, which are each connected to an \ac{apd}. All \acp{eom} are still turned off. We record the correlation histogram shown in Fig.~\ref{fig:qd_characterization}(d) and retrieve a second-order correlation function at zero time delay of $g^{(2)}(0)=0.016(1)$ by taking the ratio of the integrated counts of the center peak to the average integrated counts of the side peaks. The measured single-photon purity of our source is limited by the residual pump laser and re-excitation processes. 
Furthermore, we measure the lifetime of the neutral exciton under pulsed resonant excitation. We obtain a lifetime of \SI{167(8)}{ps} and a fine structure splitting of \SI{9.7(2)}{\micro eV} (see Sec.~S4 in the \hyperlink{supp}{supplementary material}).

From the detected count rate $R_\mathrm{det}$ at $\pi$-pulse excitation, we can calculate the fiber-coupled efficiency of our \ac{qd} single-photon source $\eta_\mathrm{QD}$, which includes all loss contributions from the source up to the routing setup. These are the finite population probability $\eta_\mathrm{pop}$, blinking of the \ac{qd} with an on-time-fraction of $\eta_\mathrm{blinking}$,\cite{Efros1997} the finite extraction efficiency $\eta_\mathrm{extr}$ from the sample, losses in the optical elements of the \ac{qd} setup with $\eta_\mathrm{optics}$, and the fiber-coupling efficiency of the \ac{qd} mode to the fiber mode $\eta_\mathrm{fibercoup}$. The estimated fiber-coupled source efficiency is
\begin{equation}
    \eta_\mathrm{QD}=\frac{R_\mathrm{det}}{RR}\;\frac{1}{\eta_\mathrm{Ch1}\eta_\mathrm{det}}=0.90(9)\%.
\end{equation}
The fiber-coupled source efficiency can also be estimated from the above mentioned loss contributions (see Sec.~S5 in the \hyperlink{supp}{supplementary material}).

Next, we characterize the efficiency of the routing setup with classical laser pulses. We connect the pump laser, which is otherwise used to excite the \ac{qd}, directly to the input of the routing setup. The continuous-wave-equivalent power of the laser is $\approx\SI{100}{\micro W}$ (measured free-space after the input coupler of the routing setup). Then, we measure the fiber-coupled power in all four output channels with a powermeter (Thorlabs PM110D and photodiode sensor S121C) and calculate the channel efficiency from the ratio of measured power divided by laser input power. We obtain an efficiency of 22.5(10)\%, 19.9(8)\%, 21.4(9)\%, and 19.9(8)\% for output Ch.~1 to 4, respectively, resulting in a combined efficiency of all channels of $\eta_\mathrm{routing}=84(3)\%$. This does not yet include a reduction in the efficiency due to erroneous switching as a consequence of the finite extinction ratio of the \acp{eom}, since we measure the power in a time-integrated fashion. We also measure the fiber-coupling efficiency of each output channel, obtained as the ratio of power measured in the fiber to power measured free-space in front of the fiber coupler. Here, we obtain values of 95(4)\%, 89(4)\%, 89(4)\%, and 87(4)\% for Ch.~1 through 4, respectively. We suspect that the variation in coupling efficiency is due to variations in the fiber coupler lens quality and in the wavefront error induced by the optical elements in the setup e.g.~the \acp{pbs}.

\begin{figure*}[t]
\includegraphics[scale=1]{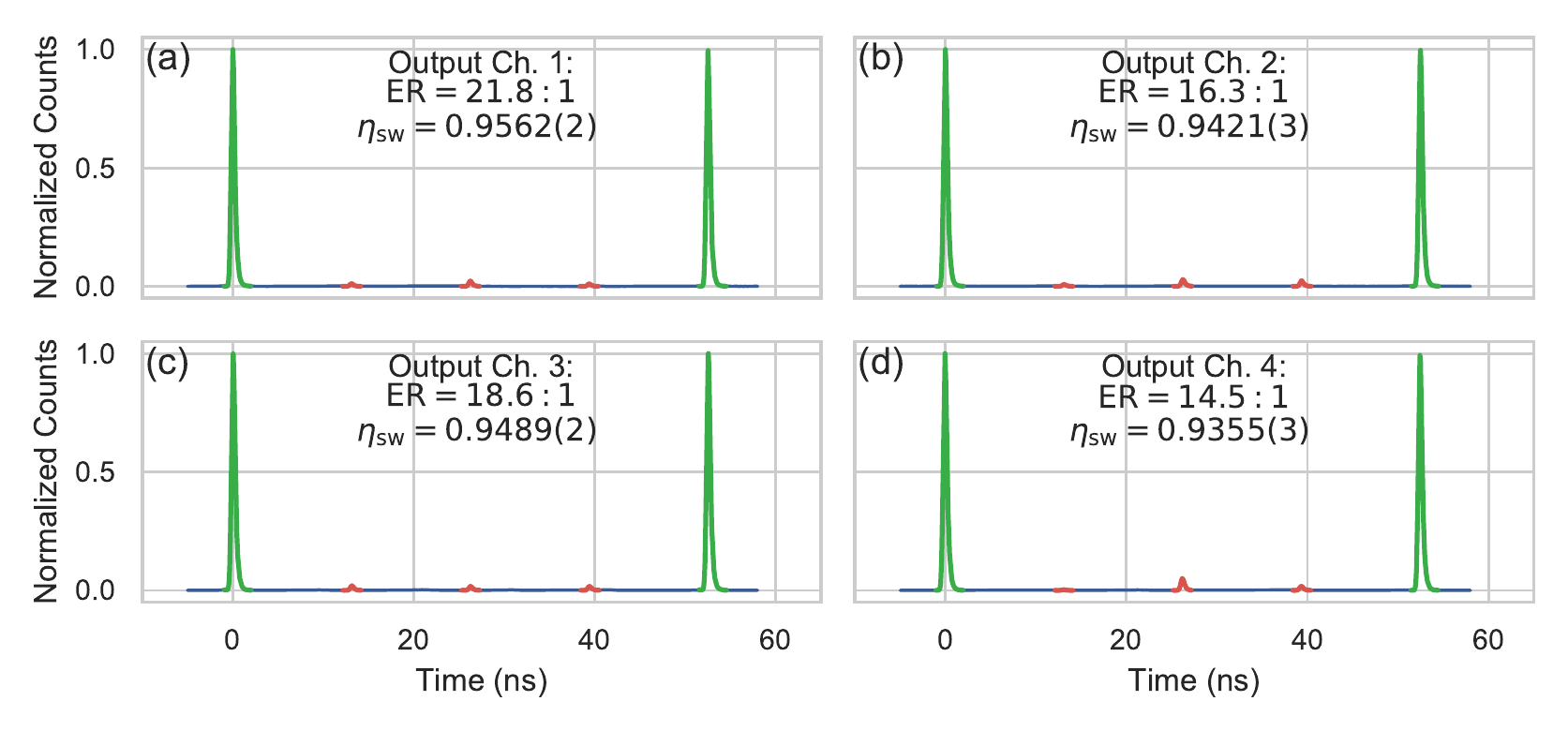}
\caption{\label{fig:extinction_ratio}Normalized histogram of accumulated time differences between the pump laser and the four output channels of the demultiplexer. Every fourth pump pulse starts the clock and the detection of a photon in an output channel stops the clock. The extinction ratio of the four output channels is calculated from the ratio of integrated counts of the main peak (green) to the sum of integrated counts of the side peaks (red). The uncertainty of the switching efficiency is calculated from the Poissonian counting statistics.}
\end{figure*}
In the next step, we characterize the switching efficiency of the routing setup $\eta_\mathrm{sw}$ given by the ratio of correctly routed events to the total number of events. The results shown in Fig.~\ref{fig:extinction_ratio} are obtained with light from the \ac{qd} single-photon source. We perform a start-stop-measurement, where the internal photodiode of the pump laser serves as a start signal (the clock is started only every fourth pulse)
and the detection of a photon in either of the four output channels stops the clock. The time differences are then accumulated into a histogram. Ideally, one should only observe peaks in the histogram separated by $4\Delta t=\SI{52.5}{ns}$. These peaks are indicated in green color in Fig.~\ref{fig:extinction_ratio}. In the experiment, we observe additional peaks marked in red that correspond to incorrectly routed events. There are three different combinations of incorrect routing which correspond to the three side peaks in Fig.~\ref{fig:extinction_ratio}. Such an incorrectly routed event could e.g., be produced in the following way: Suppose that the photon in the second time bin in Fig.~\ref{fig:setup} is incorrectly routed in transmission through the first \ac{pbs} towards \ac{eom} C. It could therefore be routed towards channel 2 or 4 and produce an event corresponding to a time difference of \SI{13.1}{ns} in Fig.~\ref{fig:extinction_ratio}(b) or \SI{39.4}{ns} in Fig.~\ref{fig:extinction_ratio}(d), respectively. There are also other erroneous routing events, which, combined, lead to the red side peaks in Fig.~\ref{fig:extinction_ratio}. The peak heights of the three side peaks are marginally different due to a slight variation of the individual \acp{er} of the \acp{eom}, a different \ac{er} of the \ac{pbs} in transmission and reflection, and the exact alignment conditions.
For each output channel, we calculate the channel \ac{er} as the ratio of the average integrated counts of the main peaks $\Sigma_\mathrm{main}$ to the sum of integrated counts of the three side peaks $\Sigma_\mathrm{side}$, as well as the switching efficiency $\eta_\mathrm{sw}=\Sigma_\mathrm{main}/(\Sigma_\mathrm{main}+\Sigma_\mathrm{side})$. The results are summarized in Fig.~\ref{fig:extinction_ratio} from which we calculate an average switching efficiency across all channels of $\eta_\mathrm{sw}=0.946(8)$, where the uncertainty is given by the standard deviation of the four single-channel switching efficiencies.
An advantage of our specific setup design is that an incorrectly routed photon will never end up in the same time bin as a correctly routed photon. A fourfold coincidence is therefore either produced by four correctly routed photons or four incorrectly routed photons, with the former being about $10^6$ times more likely.

In an active $m$-mode demultiplexer, the detected $n$-fold coincidence rate of $n$ distinct channels clicking simultaneously ($n\le m$) is given by (adapted from Ref.~\onlinecite{Lenzini2017}--we additionally take into account blinking of the \ac{qd} and differentiate between $m$ and $n$)
\begin{eqnarray}
    R(n)=&&\frac{RR}{m}\eta_\mathrm{blinking}\left(\frac{\eta_\mathrm{QD}}{\eta_\mathrm{blinking}}\eta_\mathrm{routing}\eta_\mathrm{det}\right)^n\nonumber\\
    &&\times
    \left[\eta_\mathrm{sw}^n+(m-1)\left(\frac{1-\eta_\mathrm{sw}}{m-1}\right)^n\right].
    \label{eq:coincidencerate}
\end{eqnarray}
The first and the second summand in the square bracket term corresponds to a coincidence due to $n$ correctly routed photons and the $(m-1)$ possible cases of $n$ incorrectly routed photons, respectively. A passive demultiplexer would correspond to $\eta_\mathrm{sw}=0.25$ such that the square bracket term simplifies to $m/m^n$. Blinking of the \ac{qd}--that is an \emph{on}-\emph{off}-type behavior of its photoluminensence intensity--manifests itself in a different scaling compared to the other efficiencies in Eq.~\eqref{eq:coincidencerate}. If the blinking timescale is long compared to the switching cycle time $m/RR$, then the detected coincidence rate scales linearly with the on-time-fraction $\eta_\mathrm{blinking}$ and not with $\eta_\mathrm{blinking}^n$ since on average the \ac{qd} is either \emph{on} for the complete switching cycle or \emph{off}. This is different for the other efficiencies, since they are completely uncorrelated in time. 
For the investigated \ac{qd}, we measure an on-time-fraction of $\eta_\mathrm{blinking}=0.36$ and a blinking timescale in the order of milliseconds, much longer than the switching cycle time \SI{52.5}{ns} (see Sec.~S6 in the \hyperlink{supp}{supplementary material} for a comparison of the blinking behavior without and with additional above bandgap illumination of the sample).

\begin{table*}[t]
\caption{\label{tab:rates}Measured and calculated count rates at the four output channels of the routing setup, as well as coincidence rates between all combinations of output channels.
}
\begin{ruledtabular}
\begin{tabular}{lllllllll}
 & & & & & & & Mean\footnotemark[1] & Calculated\footnotemark[2]\\
\hline
 & & & & & & & & \\
Channel & 1 & 2 & 3 & 4 & & & & \\
$R(1)$ (Hz) & \num{104e3} & \num{91e3} & \num{103e3} & \num{96e3} & & & \num{99(6)e3} & \num{98(13)e3}\\
 & & & & & & & & \\

Channel combinations  & (1,2) & (1,3) & (1,4) & (2,3) & (2,4) & (3,4) & & \\
$R(2)$ (Hz) & 1093 & 1207 & 1134 & 1044 & 981 & 1085 &  \num{1.09(8)e3} & \num{1.3(3)e3}\\
 & & & & & & & & \\
Channel combinations & (1,2,3) & (1,2,4) & (1,3,4) & (2,3,4) & & & & \\
$R(3)$ (Hz) & 14.1 & 13.3 & 14.6 & 12.7 & & & \num{14.0(9)} & 17(7)\\ 
 & & & & & & & & \\
$R(4)$ (Hz) & & & & & & & $0.17(1)$ &	0.23(12)\\

\end{tabular}
\end{ruledtabular}
\footnotetext[1]{\emph{First row:} Mean count rate averaged over all channels. \emph{Second and third row:} Mean $n$-fold coincidence rate averaged over all $n$-fold coincidence combinations. \emph{Fourth row:} Fourfold coincidence rate of all four channels clicking simultaneously. For the fourfold coincidence rate, the measurement uncertainty is obtained from the Poissonian counting statistics.}
\footnotetext[2]{Calculated according to Eq.~\eqref{eq:coincidencerate} and taking into account the estimated efficiencies of the setup.}
\end{table*}
For our four-mode demultiplexer, the measured and calculated $n$-fold coincidence rates are summarized in Tab.~\ref{tab:rates}. The measured mean $n$-fold rate agrees well with the expected $n$-fold rate, which is calculated according to Eq.~\eqref{eq:coincidencerate} and the measured efficiencies of the setup. We measure a four-photon coincidence rate of \SI{0.17(1)}{Hz} over a measurement time of $\approx\SI{15}{h}$, detecting in total $\num{8.7e3}$ fourfold coincidence events. We noticed that  $g^{(2)}(0)$ increased from initially $g^{(2)}(0)=0.016$ before the measurement to $g^{(2)}(0)=0.116$ after the measurement, most probably due to a degradation in the cross-polarized laser suppression. While this slightly increased the single count rates over the course of the measurement ($\sim 8\%$), it had almost no influence on the detected four-photon coincidence rate (see Sec.~S7 in the \hyperlink{supp}{supplementary material}). 
The device can also be used as a three-photon source by only connecting e.g.~output Ch. 1, 3, and 4, where we obtain a three-photon rate of \SI{14.6}{Hz}.

\section{\label{sec:hom}Photon indistinguishability at varying excitation pulse separations}

\begin{figure*}[t]
\includegraphics[scale=1]{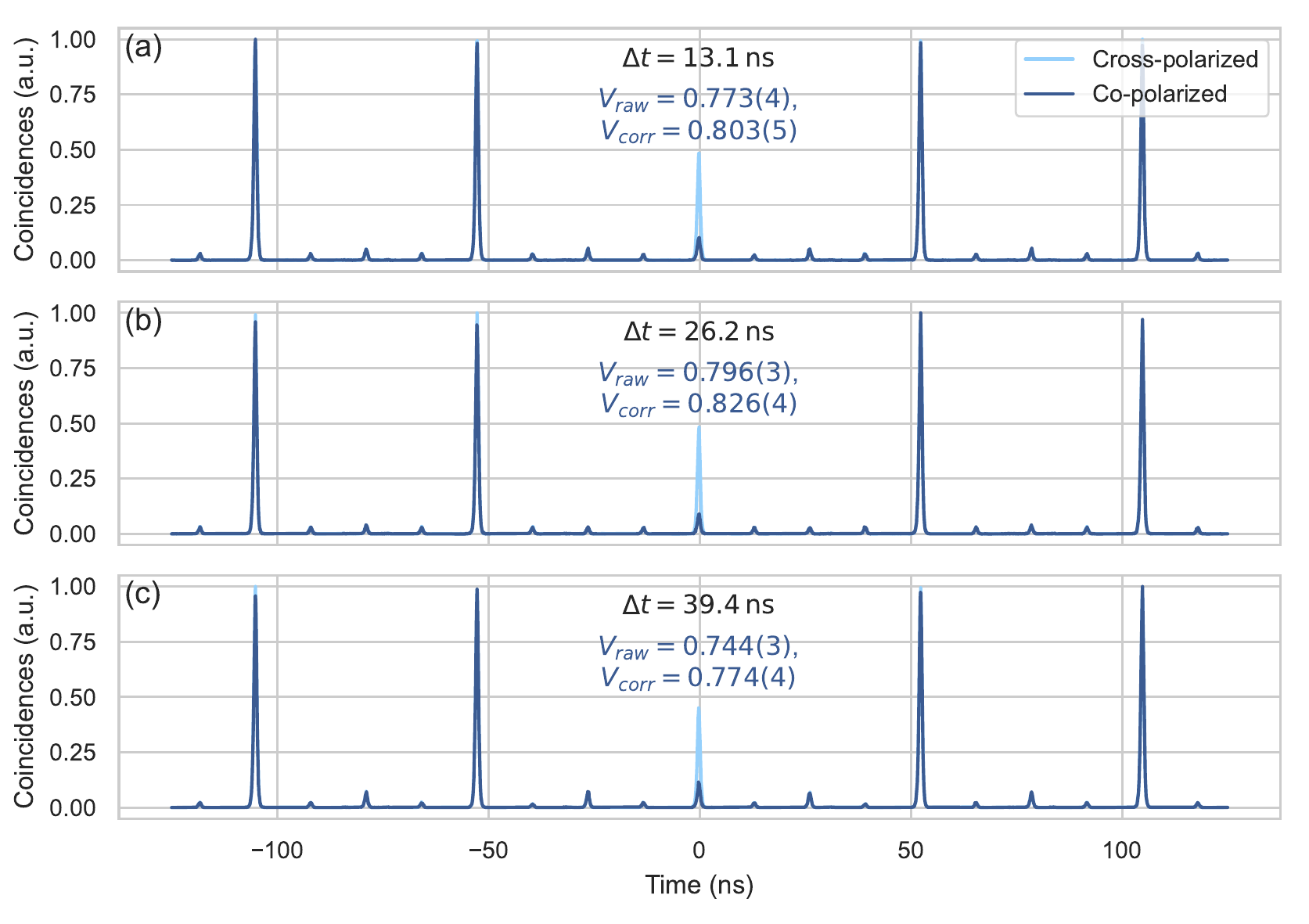}
\caption{\label{fig:hom} \acl{hom} interference measurements between different combinations of output channels from the routing setup. These combinations correspond to temporal delays between the interfering photons of (a) \SI{13.1}{ns}, (b) \SI{26.2}{ns}, and (c) \SI{39.4}{ns}. The raw \ac{hom} visibility is calculated from the integrated counts of the central peak for the co- and cross-polarized case. $V_\mathrm{corr}$ is the \ac{hom} visibility corrected for an imperfect splitting ratio of the beamsplitter as well as a finite $g^{(2)}(0)$.}
\end{figure*}
The routing setup can be used to characterize the indistinguishability of consecutively emitted photons from the \ac{qd} via \acf{hom}-type \ac{tpi}.\cite{Hong1987,Senellart2017, Ollivier2021} We perform \ac{tpi} at pump pulse separations of \num{13.1}, \num{26.2}, and \SI{39.4}{ns}, which is implemented by connecting output Ch.~1 to one input and output Ch.~2, 3, or 4 to the other input of a polarization-maintaining 50/50 fiber beamsplitter. The outputs of the beamsplitter are attached to \acp{apd}. We obtain the correlation histograms shown in Fig.~\ref{fig:hom} for a co-polarized and a cross-polarized input to the beamsplitter. Cross-polarized input is achieved by placing a \ac{hwp} oriented at \SI{45}{\degree} in front of the fiber coupler of Ch.~1. Before the measurement, we adjust the temporal overlap of the photons at the beamsplitter with the procedure described in Sec.~S8 in the \hyperlink{supp}{supplementary material}. 
We calculate the raw \ac{hom} visibility from 
\begin{equation}
    V_\mathrm{raw}=1-\frac{C_{\parallel}}{C_{\perp}},\label{eq:vraw}
\end{equation}
where $C_{\parallel}$ and $C_{\perp}$ are the normalized coincidence counts integrated over the central peak for the co-polarized and cross-polarized configuration, respectively. The three small peaks in between the central peak and the main uncorrelated peaks (uncorrelated peak e.g.~at \SI{52.5}{ns}) correspond to the cross-correlation between a correctly routed photon in one channel and an incorrectly routed photon in another channel or vice versa. 
Correcting for an imperfect splitting ratio of the beamsplitter and a non-zero $g^{(2)}(0)$, the indistinguishability is given by \cite{Ollivier2021} 
\begin{equation}
    V=\frac{V_\mathrm{raw}+g^{(2)}(0)}{1-g^{(2)}(0)}\;\frac{R^2+T^2}{2RT},
\end{equation}
with the beamsplitter intensity reflectivity (transmissivity) $R=0.514(7)$ ($T=1-R$), as well as $g^{(2)}(0)=0.016(1)$. 
We obtain corrected indistinguishabilities of $V_{\SI{13.1}{ns}}=0.803(5)$, $V_{\SI{26.2}{ns}}=0.826(4)$, and $V_{\SI{39.4}{ns}}=0.774(4)$.

The photon indistinguishability is approximately constant for the three investigated excitation pulse separations. The obtained value for \SI{39.4}{ns} is slightly smaller than for the two shorter temporal delays, but this might be caused by uncertainties in the measurement e.g.~imperfect perpendicular polarization in the cross-polarized measurement. We expect that this is the case, since the area of the center peak for the cross-polarized measurement in Fig.~\ref{fig:hom}(c) is smaller than half the area of the uncorrelated peaks. This might underestimate $V_\mathrm{raw}$ in Eq.~\eqref{eq:vraw}. If we instead calculate $V_\mathrm{raw}'=1-2A$ where $A$ is the ratio of the area of the center peak to the average area of the uncorrelated peaks, then we obtain a value of $V'_{\SI{39.4}{ns}}=0.795$ very close to the values obtained for the other two temporal delays.

A degradation in the indistinguishability for larger excitation pulse separation is expected due to charge fluctuations in the vicinity of the \ac{qd}. Charge-noise causes spectral diffusion and an inhomogeneous broadening of the \ac{qd} transition.\cite{Thoma2016} It has been shown that a degradation is already present at short timescales for the GaAs \acp{qd} studied here and excited via two-photon excitation ($V=0.62$ at \SI{2}{ns} and $V=0.43$ at \SI{12.5}{ns} pulse separation),\cite{Huber2017} as well as LO-phonon excitation ($V=0.92$ at \SI{4}{ns} and $V=0.78$ at \SI{12}{ns}),\cite{Reindl2019} and in InGaAs \acp{qd} excited quasi-resonantly ($V=0.94$ at \SI{2}{ns} and $V=0.53$ at \SI{12.5}{ns}).\cite{Thoma2016}
The above findings are confirmed by photon-correlation Fourier spectroscopy, which directly probes the timescale of spectral diffusion of the \ac{qd} transition.\cite{Schimpf2019} From these examples we conclude that charge-noise is the main cause of the imperfect indistinguishability observed in our measurement, given the large pulse separation times in our measurement. Furthermore it has been shown for InGaAs \acp{qd}, that for an even larger excitation pulse separations a plateau-like behavior is theoretically expected.\cite{Thoma2016} 
Our findings exhibiting a constant indistinguishability from \num{13.1} to \SI{39.4}{ns} pulse separation
indicate that such a plateau-like behavior can also be reached in the GaAs material system.
Charge-noise can be strongly reduced by improved material quality and by embedding the \acp{qd} in charge-tunable devices as already demonstrated for InGaAs\cite{Somaschi2016,Tomm2021} and GaAs\cite{Zhai2021} \acp{qd}.

An additional mechanism possibly leading to a deterioration of photon indistinguishability is phonon-induced pure dephasing. In Ref.~\onlinecite{Thoma2016}, the authors conclude that for InGaAs \acp{qd} a degradation in the photon indistinguishability is almost negligible for temperatures $<\SI{10}{K}$, but has a significant effect at higher temperatures. 
In our experiment we see that the \ac{tpi} visibility keeps increasing when lowering the temperature from nominal \SI{9}{K} to \SI{8}{K} (see Sec.~S9 in the \hyperlink{supp}{supplementary material}), indicating that the \ac{qd} studied here may be particularly sensitive to phonon-induced pure dephasing. This finding is in line with the fact that the \ac{qd} emits at a wavelength of almost \SI{800}{nm}, which implies a large \ac{qd} with small spacing among confined levels and consequently enhanced zero-phonon-line broadening due to virtual transitions to excited states.\cite{Muljarov2004} 

The indistinguishability can potentially be improved by narrow-band spectral filtering which suppresses the phonon-sidebands at the cost of overall efficiency. Similarly, this can be realized by embedding the \ac{qd} layer inside a narrow-band cavity.\cite{Wang2016a}


\section{\label{sec:conclusion}Comparison and conclusion}

\begin{table*}[t]
\caption{\label{tab:comparison}
Comparison of this work to other temporal-to-spatial demultiplexing  implementations.
}
\begin{ruledtabular}
\begin{tabular}{lllllll}
Reference & Photon source & Demultiplexer & switching & $\eta_{\mathrm{QD}}$ & $\eta_\mathrm{routing}\eta_\mathrm{sw}$ & measured $R(4)$\\
 & (excitation scheme) & implementation & rate & $R_{\mathrm{det}}$\footnotemark[1] & & \\
 &  & & & $R_\mathrm{QD}$\footnotemark[2] &  & \\
\hline
  &  &  &  &  &  & \\
\onlinecite{Wang2019} & InGaAs QD in micropillar & Free-space,& \SI{0.76}{MHz} & 26.1\% & 84\%\footnotemark[3] & $\approx\SI{3000}{Hz}$\footnotemark[4]\\
  & (resonant excitation) & broadband \acp{eom} & & \SI{16.3}{MHz} &  & \\
  &  &  &  & \SI{19.9}{MHz} &  &  \\
  &  &  &  &  &  & \\
\onlinecite{Hummel2019a}, \onlinecite{Hummel2019} & InGaAs QD in & Free-space, & \SI{0.95}{MHz} & 2.8\% & 76.7\% & $\SI{1.05}{Hz}$\\
  & integrated waveguide & broadband \acp{eom} &  & \SI{1.7}{MHz} & ($\eta_{fiber}\eta_{sw}\eta_{m}$)\footnotemark[5] & \\
  & (non-resonant excitation) & & & \SI{1.9}{MHz} &  & \\
  &  &  &  &  &  & \\
\onlinecite{Lenzini2017} & InGaAs QD in micropillar & Integrated switches & \SI{40}{MHz} & $\approx 2.5\%$ & 23\% & \SI{0.18}{mHz}\\
  &  (quasi-resonant excitation) & in lithium niobate & & & ($\eta_\mathrm{DM}T$)\footnotemark[5] & (estimated)\\
  &  &  &  &  &  & \\
  &  &  &  &  &  & \\
  \onlinecite{Pont2022} & InGaAs QD in micropillar & Acousto-optic modulator & \SI{1.4}{MHz} & $\approx 9.5\%$ & $\approx 65\%$ & \SI{1.6}{Hz}\footnotemark[4]\\
  & (LA-phonon excitation) & &  &  &  & \\
  &  &  &  &  &  & \\
This work & GaAs QD in planar cavity & Free-space, resonantly & \SI{38.1}{MHz} & 0.90(9)\% & 79(3)\% & \SI{0.17(1)}{Hz}\\
 & (resonant excitation) & enhanced EOMs & & \SI{0.425(4)}{MHz} &  & \\
 &  &  &  & \SI{0.69(7)}{MHz} &  & \\
\end{tabular}
\end{ruledtabular}
\footnotetext[1]{Detected count rate of the \ac{qd} single-photon source.}
\footnotetext[2]{Count rate of the \ac{qd} single-photon source corrected for detection efficiency.}
\footnotetext[3]{It is not mentioned in Ref.~\onlinecite{Wang2019} whether this value includes or excludes the switching efficiency $\eta_\mathrm{sw}$ of the setup.}
\footnotetext[4]{After low loss linear optical network.}
\footnotetext[5]{Variable names as given in the reference.}
\end{table*}
We demonstrated active temporal-to-spatial demultiplexing by utilizing, for the first time, resonantly enhanced free-space \acp{eom}. In Table~\ref{tab:comparison}, we compare our work to other active temporal-to-spatial demultiplexing implementations. All other implementations so far have been performed with InGaAs \acp{qd}\cite{Wang2019,Hummel2019a,Lenzini2017} and the routing setup was either implemented with broadband free-space \acp{eom}\cite{Wang2019,Hummel2019a}, integrated switches (with a low overall efficiency),\cite{Lenzini2017} or acousto-optic modulators. Compared to the broadband counterpart, resonantly enhanced \acp{eom} require much lower half-wave voltages and therefore no bulky and expensive high-voltage amplifiers to drive the \ac{eom}. In addition, due to the lower half-wave voltage, these \acp{eom} can be operated at much higher switching rates (in our case \SI{38.1}{MHz}). This is beneficial, since the emission time separation between temporally overlapping photons after the demultiplexer is much shorter. In this work, the emission time separation is 13.1 to $\SI{39.4}{ns}$ compared to $\approx\SI{1}{\micro s}$ in Ref.~\onlinecite{Wang2019,Hummel2019a, Pont2022}. 
As discussed in Section \ref{sec:hom}, our demultiplexing implementation improves the indistinguishability of the photons by reducing the influence of charge-noise.
For example in Ref.~\onlinecite{Pont2022}, the \ac{tpi} visibility decreased from 92\% at \SI{12}{ns} to 76\% at \SI{960}{ns} emission time separation.
In addition, a high switching rate is also advantageous in multi-photon interference experiments with current single-photon detection technology that relies on \acp{apd} and \acp{snspd}. These detectors have typical dead times in the order of tens of nanoseconds, which matches well with the switching cycle time of \SI{52.5}{ns} of our demultiplexer. In contrast, in a demultiplexer with broadband \acp{eom} operated at $\approx\SI{1}{MHz}$, the photons arrive in bunches of about 20 narrowly spaced photons and, therefore, the detection of a photon in the first time bin might render the detector unable to detect a photon in the subsequent time bins due to detector dead time. This might not be an issue for a lossy setup which makes it very unlikely that two consecutive time bins are both occupied with a photon, but becomes an important factor in the case of close to unity overall efficiency.

Our implementation of a demultiplexer results in an overall routing setup efficiency of $\eta_\mathrm{routing} \eta_\mathrm{sw}=79(3)\%$ comparable to the implementations with broadband free-space \acp{eom}\cite{Wang2019,Hummel2019a} and much better than the integrated implementation.\cite{Lenzini2017} We measure lower four-photon rates compared to Ref.~\onlinecite{Wang2019, Hummel2019} as a result of our much lower source brightness. With a source brightness of 26.1\% as in Ref.~\onlinecite{Wang2019}, we predict a detected four-photon rate of \SI{7(2)e3}{Hz} with our demultiplexer. It is therefore crucial to increase the source brightness and reduce the overall losses in the setup as one can likewise infer from Eq.~\eqref{eq:coincidencerate}. The source brightness could be increased by embedding the GaAs \ac{qd} into a micropillar or circular Bragg grating structure which should greatly improve the extraction efficiency.\cite{Ding2016,Somaschi2016,Senellart2017,Liu2019}

The setup can be easily scaled up to a larger number of $2^k$ ($k$ being an integer) output channels e.g.~8 or 16. This is done by recursively adding an additional layer of four, eight or in general $2^{k-1}$ \acp{eom} with a switching rate of $RR/8$, $RR/16$ or in general $RR/2^k$ to the previous setup. Additionally, due to the short decay time of the \ac{qd} transition (typically 150-\SI{250}{ps} without resorting to Purcell enhancement), one could operate the single-photon source also at a higher pump rate of e.g.~\SI{152.4}{MHz} (obtainable by passive laser-pulse multiplexing), which would require an additional \ac{eom} operating at \SI{76.2}{MHz}.\footnote{A factor of two faster \acp{eom} should be feasible according to the manufacturer.} This would increase the obtainable multiphoton rate by a factor of two.
Moreover, the demultiplexer can be used with any other single-photon source operated at a similar repetition rate and emission wavelength.

Active temporal-to-spatial mode demultiplexing of a \ac{qd} single-photon source paves the way for truly deterministic multi-photon sources. This provides a feasible route towards multi-photon interference experiments with a large and precisely defined number of photons, which, so far, has been inaccessible with \ac{spdc} photon sources due to the probabilistic nature of their emission. 

\section*{Supplementary material}

\hypertarget{supp}{}
See the \hyperlink{supp}{supplementary material} for additional information on the sample structure, the \ac{eom}-\ac{pbs} transmission function, the damped rabi oscillation model (fitting function in Fig.~\ref{fig:qd_characterization}(b)), the lifetime of the \ac{qd} emission, an overview of the \ac{qd} setup efficiencies, an analysis of \ac{qd} blinking, the timetrace of detected count rates, the interferometric path length adjustment procedure, and the temperature dependence of the \ac{tpi} visibility.

\begin{acknowledgments}
The authors acknowledge assistance from M.~Aigner and fruitful discussions with M.~Reindl, D.~Huber, C.~Schimpf and E.~Vogt (QUBIG GmbH).

This work was financially supported by the Austrian Science Fund (FWF) via the projects P 30459 and I 4320 with D.~Reiter, the Research Group FG5, the European Union’s Horizon 2020 research and innovation program under Grant Agreements No.~899814 (Qurope) and No.~871130 (Ascent+), the Linz Institute of Technology (LIT), and the LIT Secure and Correct Systems Lab, supported by the State of Upper Austria.
\end{acknowledgments}

\section*{Author Declarations}

\subsection*{Conflict of Interest}

The authors have no conflicts to disclose.

\section*{Data Availability Statement}

The data that support the findings of this study are openly available in Zenodo at \hyperlink{http://doi.org/10.5281/zenodo.6337697}{http://doi.org/10.5281/zenodo.6337697}. 

\FloatBarrier 

\section*{References}

\bibliography{ms}

\end{document}